\documentclass[pre,showpacs,amsmath,amssymb]{revtex4} 
\usepackage{epsf}
\usepackage{graphicx}
\usepackage{dcolumn}
\usepackage{bm}
 
\begin{document}
 
\title{Asymptotic Dynamics of Breathers in Fermi-Pasta-Ulam Chains}
\author{R. Reigada\footnote{Permanent address:
Departament de Qu\'{\i}mica F\'{\i}sica,
Universitat de Barcelona, Avda. Diagonal 647, 08028 Barcelona, Spain},
A Sarmiento\footnote{Permanent address:
Instituto de Matem\'aticas, Universidad Nacional
Aut\'onoma de M\'exico, Ave. Universidad s/n, 62200 Chamilpa,
Morelos, M\'exico}, Katja Lindenberg}
\affiliation{Department of Chemistry and Biochemistry and Institute for
Nonlinear Science, University of California San Diego, 9500 Gilman Drive,
La Jolla, CA 92093-0340}
\date{\today}

\begin{abstract}
We study the asymptotic dynamics of breathers in finite Fermi-Pasta-Ulam
chains at zero and non-zero temperatures.  While such breathers are
essentially stationary and very long-lived at zero temperature, thermal
fluctuations tend to lead to breather motion and more rapid decay.
\end{abstract}

 
\maketitle
\section{Introduction}
\label{introduction}
Energy localization in the form of breathers has been intensely
investigated over the past several years \cite{various}. 
These highly localized long-lived
excitations in translationally invariant nonlinear arrays are of great
interest because they provide a mechanism for energy storage that does not
rely on defects.  Indeed, the fact that these excitations are often mobile
(which they would in general not be if the localization were
defect-induced) makes them particularly interesting in the context
of efficient transport of vibrational energy~\cite{directed}. 

Exact breather excitations, that is, breathers that persist forever, can be
confirmed analytically and constructed numerically for certain nonlinear
infinite arrays of interacting masses, among them nonlinear arrays with
no local potentials and interactions between neighboring sites
that vary as $(x_i-x_j)^n$ with $n\rightarrow \infty$ \cite{breather}.
The $x$'s
denote displacements of the masses at sites $i$ and $j$.  When the
interactions are not precisely of this form, or when the array is finite,
or when other excitations (for example, phonons or other
localized excitations) are present in the system, it is no longer possible
to prove that there are exact breather solutions.
Nevertheless, it is possible to explore the problem numerically.
In particular, in this paper we study the asymptotic dynamics of
localized excitations in the one-dimensional Fermi-Pasta-Ulam (FPU) 
$\beta$-model at zero temperature and during thermal relaxation.
We corroborate that long-wavelength thermal excitations have a
profound effect on breather stability.  Since thermal
fluctuations are inevitable in any real system, zero-temperature
results must be applied with appropriate caution.
These results extend our earlier
work to much longer time regimes~\cite{our1d,our2d}, and agree with
and complement those of Piazza et al.~\cite{piazza1}. The conditions
for breather stationarity and longevity are thereby further clarified.
We present our numerical results in Secs.~\ref{zero}
and \ref{nonzero}, and end with a brief summary in Sec.~\ref{summation}.

\section{Breathers at Zero Temperature}
\label{zero}
The Hamiltonian for the FPU $\beta$-model is
\begin{equation}
H = \sum_{i=1}^{N} \frac{\dot{x}_{i}^2}{2} + \frac{k}{2} \sum_{i=1}^{N}
(x_{i} - x_{i-1})^2 +\frac{k'}{4} \sum_{i=1}^{N} (x_{i} - x_{i-1})^4,
\label{ham1}
\end{equation}
where $N$ is the number of sites, and $k$ and $k'$ are the harmonic
and anharmonic force constants, respectively. We set
$k=k^\prime=0.5$ throughout.  The equations of motion 
with free-end boundary conditions are integrated using a
fourth order Runge-Kutta method with time interval $\Delta t=5\times
10^{-4}$ (further reduction leads to no significant improvement).  
The total energy of the array 
is the sum over individual symmetrized site energies $E_i(t)$~\cite{our1d}.
A zero temperature environment with the least disturbance
to the dynamics in the chain is achieved by connecting only
the ends of the chain to such an environment via a
dissipation term.  This is accomplished by adding a term of the form
$-\gamma \dot{x}_i$ to the equations of motion for sites $i=1$ and
$i=N$~\cite{piazza1,our1d}. 

Since we do not {\em a priori} know the form of the longest-lived localized
excitation, we initially create an excitation of the ``odd parity" breather
configuration with amplitudes
$(0,\cdots,0,-A/2,A,-A/2,0,\cdots,0)$, and zero velocity at each site. The
frequency of the resulting oscillatory motion increases with increasing
$A$, which must be chosen so that this frequency is
above the phonon band edge at $\omega=\sqrt{2}$.  This
excitation is not an exact
stationary solution of the equations of motion, so it typically sheds
some energy
while re-accommodating amplitudes and frequencies, and settles into an
excitation that is very long-lived.  The discarded energy 
appears in the form of phonons that travel away from the localized
excitation and
dissipate quickly, in a time $\tau_m$, across
the ends of the chain.  
The balance of the energy remains localized, most of it ($98\%$ for
$A=0.5$) on the three initially excited sites, and decays exponentially
with an extremely long characteristic time $\tau$. 

\begin{figure}[htb]
\begin{center}
\epsfxsize = 3.0in
\epsffile{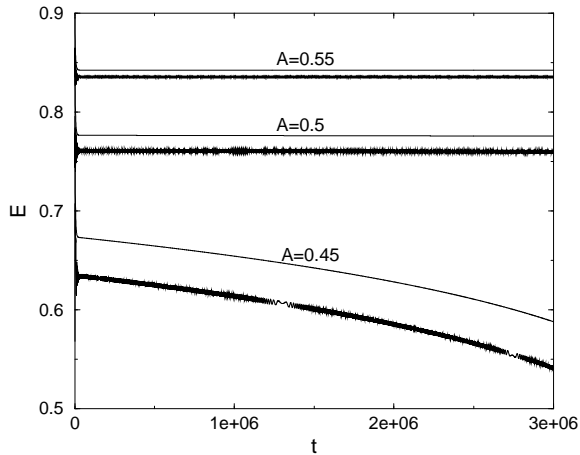}
\end{center}
\caption{Decay of the normalized energy $E(t)$ for three different values
of the initial amplitude $A$ for chains of $N=31$ sites connected at the
ends to a zero-temperature bath.  The dissipation parameter $\gamma=1$. The
thin lines represent the total energy remaining in the chain, and the bold
lines the portion of the remaining energy that is localized on the three
initially excited sites.
}
\label{fig1}
\end{figure}

To extract exponents and characteristic times, it is 
crucial to have data that covers both time scales,
and to normalize the data in such a way
that one behavior does not mask or distort the other~\cite{intpiazza}.
We introduce the {\em normalized energy} $E(t)$
and the {\em modified normalized energy} $E_m(t)$:
\begin{equation}
E(t)\equiv \frac{\sum_{i=1}^N E_i(t)}{\sum_{i=1}^N E_i(0)}, \qquad
E_m(t)\equiv \frac{\sum_{i=1}^N E_i(t)}{\sum_{i=1}^N E_i(\tau_m)}.
\label{energy}
\end{equation}
The denominator in the first contains the initial energies, 
and in the second the energies after the 
discarded phonons have dissipated (but before the remaining localized
excitation has decayed appreciably). In our illustrations we take
$\tau_m=40,000$. 

\begin{figure}[htb]
\begin{center}
\epsfxsize = 3.0in
\epsffile{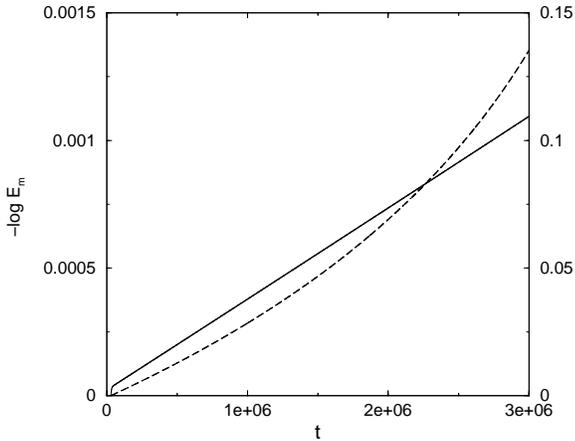}
\end{center}
\caption{$\left[-\log E_m(t)\right]$ vs $t$ for two initial
amplitudes, $A=0.5$ (solid curve, left scale)
and $A=0.45$ (dashed curve, right scale) for a chain of $31$ sites. 
}
\label{fig2}
\end{figure}

In Fig.~\ref{fig1} we show typical results for the normalized energy
for three excitation amplitudes over more than six decades of time.
The modified normalized energy shows essentially the same behavior.
If the decay of the long-lived excitation is exponential, we expect
$\left[-\log E(t)\right]$ and $\left[-\log E_m(t)\right]$
vs $t$ to be straight 
lines over the appropriate long time intervals. In Fig.~\ref{fig2} we 
clearly see this behavior, which extends over the entire time interval for
the higher (but not the lower)
amplitude excitation.  The slope for the $A=0.5$ curve
leads to a decay time of $\tau=5.32\times 10^{9}$, a number
reported here merely to stress its enormous magnitude. 
The frequency of the localized breather
increases with increasing amplitude and hence is more separated
from the phonon band for larger $A$,  leading to a slower decay
of more energetic breathers.  As the breather loses its energy, its
frequency decreases, and eventually as it approaches the phonon band
edge the decay becomes more rapid. This behavior is evident in the $A=0.45$
curve in Fig.~\ref{fig2}.  We also find that longer chains exhibit a longer
decay time: residual amplitude of the breather at sites far from
its center decreases rapidly with increasing distance and thus 
the chain end sites have less energy to dissipate. 

\begin{figure}[htb]
\begin{center}
\epsfxsize = 3.0in
\epsffile{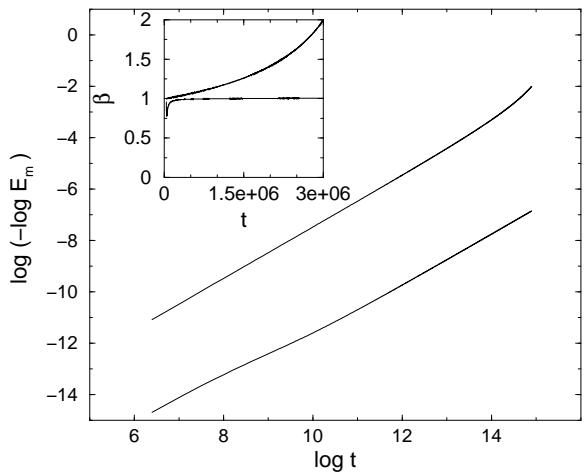}
\end{center}
\caption{$\left[\log(-\log E_m(t))\right]$ vs
$\log t$ for two initial amplitudes,
$A=0.5$ (lower curve) and $A=0.45$ (upper curve) for a chain of $31$
sites.  The inset shows the associated slopes. Lower curve:
$A=0.5$; upper curve: $A=0.45$.
}
\label{fig3}
\end{figure}

Figure~\ref{fig3} confirms the exponential 
behavior of the breather decay~\cite{piazza1,intpiazza}. 
For two initial amplitudes, the figure shows 
$\left[\log(-\log E_m(t))\right]$ vs $\log t$, which 
yields a straight
line of slope $\beta$ if the decay is of the form $E_m(t) \propto
e^{(-t/\tau)^\beta}$.  The inset shows the values of the slopes as a function
of time. For $A=0.5$ 
pure exponential behavior $\beta=1$ is confirmed throughout the time range
presented.  The deviation from pure exponential
decay for the lower amplitude breather is evident. 

\begin{figure}[htb]
\begin{center}
\epsfxsize = 3.0in
\epsffile{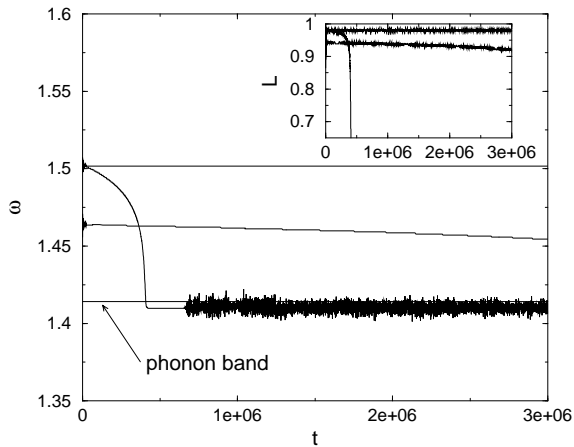}
\end{center}
\caption{Breather frequency as a function of time.  Upper curve:
$A=0.5$, $N=31$. Middle curve: $A=0.45$, $N=31$. Lower curve: $A=0.5$,
$N=21$.  Inset: associated localization parameters in the same order.
}
\label{fig4}
\end{figure}

A breather of a given amplitude has a well-defined characteristic
frequency.  In Fig.~\ref{fig4} we show this frequency in relation to the
phonon band edge as a function of time for the cases discussed above. 
For 31-site chains, the 
frequency of the breather of initial amplitude $A=0.5$ decreases very
little
over the entire simulation, while that of initial amplitude
$A=0.45$ decreases more markedly.  Consistent with the fact that the breather
does not disappear entirely in the time range shown,
its frequency never reaches
the phonon band edge.  If the initial amplitude of the excitation were even
smaller, or the simulation time much longer, or the chain shorter, the
breather would eventually disappear. This last case is illustrated in
Fig.~\ref{fig4}.  The disappearance of the
breather coincides with its frequency reaching the band edge.  The inset
shows $L$, the ratio of the energy on the five sites around the
breather to the total energy.  This parameter is of order unity when
most of the energy is localized on a small number of sites.

\section{Breathers at Finite Temperatures}
\label{nonzero}
The situation is less certain when a localized excitation evolves in a
thermal environment.  Other excitations in the medium perturb the breather,
and its eventual fate varies from one realization to another.  To
illustrate the complexity of the situation, we present two scenarios.

\begin{figure}[htb]
\begin{center}
\epsfxsize = 3.0in
\epsffile{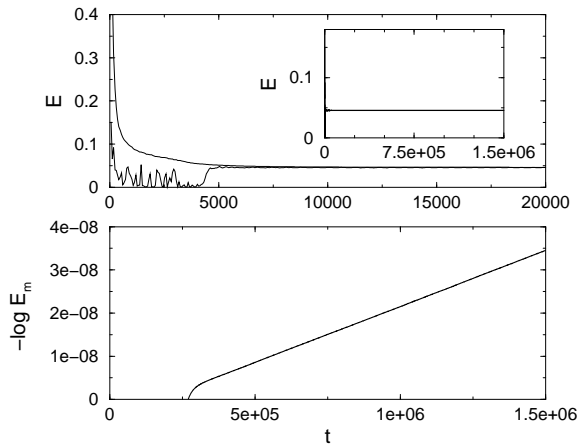}
\end{center}
\caption{Upper panel: the smooth curve is the normalized energy as a
function of time for a chain of $30$ sites initially thermalized at $T=0.5$
and connected through its ends to a zero-temperature heat bath
($\gamma=0.1$).  The initially jagged curve is the
normalized energy on sites $13$, $14$, $15$, and $16$. The inset shows the
temporal evolution of the energy on these four sites
over a longer time scale. Lower panel:  
$\left[-\log E_m(t)\right]$ vs $t$ for the same chain.
}
\label{fig5}
\end{figure}

We begin with a chain that is initially in thermal equilibrium at a nonzero
temperature $T\neq 0$~\cite{our1d}.
At time $t=0$,
the end sites of the thermalized chain are connected to a zero
temperature bath via a purely
dissipative term~\cite{our1d,piazza1,tsironic}. 
We follow the thermal relaxation of the chain and, in
particular, the dynamics of breathers that may spontaneously appear during
relaxation if the initial temperature is sufficiently high. 

Figure~\ref{fig5} shows the evolution of the total energy in a
chain of $30$ sites initially thermalized at temperature $T=0.5$.
The figure also shows the energy
in only the four sites $i=13, 14, 15, 16$.  After a relatively
short time ($5000$ time units in this particular realization) almost all
of the energy settles in these sites and remains there.
The excitation around the four sites turns out
to be an ``even parity" breather, with maximum displacements $A$ and $-A$
alternating on sites $14$ and $15$, smaller but not negligible amplitudes
at sites $13$ and $16$, and essentially no motion of the other sites.
The frequency of the breather, initially $\omega=1.633$,
decreases very little for the duration of the simulation.  In
Fig.~\ref{fig5} we also present the modified normalized energy.
Its decay is clearly exponential, with
an enormously long time constant, $\tau=4\times 10^{13}$. 
Thus this breather, even in our relatively short chain, is
essentially stationary.  This decay
time is much longer than that reported in the previous section, consistent
with the fact that the amplitude of the breather that has emerged
spontaneously is larger. 
The thermal relaxation process has swept the lattice clean of
all other excitations, allowing the breather to survive
undisturbed. This
process was described in our earlier work~\cite{our1d,our2d}.
There we reported a 
stretched exponential rather than a purely exponential decay, a
conclusion that relied on the normalized rather than the modified
normalized energy~\cite{intpiazza}, on a time history that was
not sufficiently long, and on a statistical average over a 
thermal ensemble that included realizations
where no breather appeared during the thermal relaxation process
(and where the energy dissipation was consequently much
more rapid).

In our second scenario, at $t=0$ a breather is explicitly injected
into a chain that is in thermal equilibrium at a very
low temperature (too low for the
spontaneous formation of breathers).  The 
chain is then allowed to relax into a zero
temperature heat bath.  We find that the thermal background invariably 
sets the breather in motion, and causes the breather to collide with other
excitations and with the chain boundaries. Collision events not only cause
the breather to keep moving, but also lead to breather degradation through
loss of energy upon collision~\cite{our1d,our2d}.  The resulting
lifetime $\tau$ of the breather is then in general much smaller than
in the first scenario. The lifetime is also much smaller
than that of a breather of the same amplitude injected into
a zero temperature chain.  We find this behavior
even when the temperature is extremely low (all the way
down to $T=10^{-7}$).  As a quantitative check we have
explicitly injected a breather of initial amplitude $A=0.5$ into a
chain thermalized at $T=10^{-6}$.  The system is then allowed to relax into
a zero temperature bath, as before.  Compared to the lifetime of the
breather in the zero temperature chain, the lifetime of the breather is now
much reduced, $\tau=1.3\times 10^6$. A similar comparison with $A=0.55$
again leads to dramatically different lifetimes, $\tau=1.426 \times 10^{11}$
($T=0$) and $\tau=2.1\times 10^6$ ($T=10^{-6}$).
Note that it does not much
matter whether the injected breather is of even or odd parity (we
inject an odd one).  

Why is a breather created spontaneously during thermal relaxation
more stable than one explicitly injected into a thermalized chain,
even at extremely low
temperatures? The answer lies in the effect of different phonons on
breather dynamics~\cite{piazza1,our1d,our2d}.  Whereas short wavelength
zone-boundary phonons contribute to spontaneous breather formation
(``modulational instability"), breathers are
most strongly perturbed by the longest wavelength phonons. These are also
the phonons that dissipate most rapidly out of a chain with free-end
boundaries into a zero-temperature bath.  In the higher temperature system,
when the breather is created spontaneously the long wavelength phonons have
already dissipated and, as dissipation continues up the phonon spectrum, the
breather is increasingly less disturbed until it reaches a spatially
stationary very long-lived configuration~\cite{our1d}.  On the other hand,
if the breather is injected into a thermalized system, the 
breather is subject to strong disturbance by long wavelength phonons
even at the very lowest temperatures until these phonons dissipate. An
injected breather in the thermalized scenario is therefore a more fragile
excitation than a spontaneously created breather of the same amplitude.
To confirm this description we have followed the dynamics of a
breather injected into a relaxing chain {\em after} the long
wavelength phonons
have decayed, and find the breather to be almost as stable as one in a
zero-temperature simulation.  For example, for a zero-temperature injected
breather of initial amplitude $A=0.6$ in a chain of 31 sites
($\omega=1.65$), we find a lifetime $\tau=3.10\times10^{14}$. In a chain
initially thermalized at $T=10^{-5}$ and then allowed to relax, if we wait
until $t=15,000$ before injecting the same breather we find a somewhat
shortened but still very long
lifetime of $\tau=6.15\times10^{13}$, in any case much longer than it would
be if injected at $t=0$.  We have also observed a breather in a
chain whose temperature is maintained at an
extremely low but nonzero value.  The breather in this case is always
fragile, continuing to move and lose energy until it degrades completely.
The zero temperature stability of the breather is therefore a somewhat
singular result.

\section{Summation}
\label{summation}
Our work supports three main conclusions concerning the dynamics of
breathers in FPU $\beta$-chains:
1) At zero temperature in finite chains, breather-like
excitations remain stationary
and localized, and their energy decays exponentially with
time; 2) The decay time at zero temperature is extremely
long, increasing with increasing chain length and with increasing
initial breather
amplitude; 3) Long wavelength thermal background  
sets the breather in motion, which in turn leads to a more
rapid decay of the breather.  The zero-temperature result is in this
sense fragile.

We have confirmed that breather decay for any particular realization is
exponential~\cite{piazza1}  until the breather frequency closely
approaches the phonon band edge.  However, since
the characteristic exponential time varies considerably from one
realization to another, an ensemble averaged decay may be of stretched
exponential or other form.

\section*{Acknowledgments}
The authors are grateful to F. Piazza for enlightening discussions.
This work was supported by the Engineering Research Program of
the Office of Basic Energy Sciences at the U. S. Department of Energy
under Grant No. DE-FG03-86ER13606. Support was also
provided by a grant from the University of California Institute for
M\'exico and the United States (UC MEXUS) and the Consejo Nacional de
Ciencia y Tecnolog\'{\i}a de M\'{e}xico (CONACYT).

\end{document}